\documentclass[12pt]{article}
\usepackage{amsmath}
\usepackage{amssymb}
\usepackage{amsfonts}

\DeclareMathOperator{\sech}{sech}
\newcommand{\field}[1]{\mathbb{#1}}
\newcommand{\R}{\field{R}}
\newcommand{\C}{\field{C}}

\title{
Comment on `Supersymmetry, PT-symmetry and spectral bifurcation'}
\author{B Bagchi$^a$, C Quesne$^b$\\ 
{\small $^a$ {\sl Department of Applied Mathematics, University of Calcutta,}} \\ 
{\small {\sl 92 Acharya Prafulla Chandra Road, Kolkata 700 009, India}}\\ 
{\small $^b$ {\sl Physique Nucl\'eaireTh\'eorique et Physique Math\'ematique,  Universit\'e Libre de Bruxelles,}} \\ 
{\small {\sl Campus de la Plaine CP229, Boulevard~du Triomphe, B-1050 Brussels, Belgium}}\\
{\small {\sl E-mail: bbagchi123@rediffmail.com and cquesne@ulb.ac.be}}}
\date{ }
\begin{document}
\baselineskip=20pt plus 1pt minus 1pt
\maketitle

\begin{abstract}
We demonstrate that the recent paper by Abhinav and Panigrahi entitled `Supersymmetry, PT-symmetry and spectral bifurcation' [Ann.\ Phys.\ 325 (2010) 1198], which considers two different types of superpotentials for the PT-symmetric complexified Scarf II potential, fails to take into account the invariance under the exchange of its coupling parameters. As a result, they miss the important point that for unbroken PT-symmetry this potential indeed has two series of real energy eigenvalues, to which one can associate two different superpotentials. This fact was first pointed out by the present authors during the study of complex potentials having a complex $sl(2)$ potential algebra. 
\end{abstract}

\vspace{0.5cm}



%
%
\newpage

Ever since the appearance of the seminal work of Bender and Boettcher in 1998 \cite{bender98}, PT-symmetric quantum mechanics continues to remain a prominent discipline of enquiry \cite{bender07}. These authors conjectured that the whole class of Schr\"odinger Hamiltonians invariant under the joint action of parity (P) and time reversal (T) may exhibit real or complex conjugate pairs of energy eigenvalues under some conditions related to PT being exact or spontaneously broken. Among the numerous directions of research pursued, a particularly interesting one that deserves mention is the implementation of complex $sl(2)$ as a potential algebra for the Schr\"odinger equation \cite{bagchi00, bagchi02a} and its subsequent connection \cite{bagchi01, bagchi02b} with the extension of supersymmetric quantum mechanics to non-Hermitian Hamiltonians. In particular, for the PT-symmetric complexified Scarf II potential (PCS) it was found, in the framework of two non-commuting inter-connecting complex $sl(2)$, that for unbroken PT-symmetry there are in general two series of energy levels associated with it. In this regard the conventional Hermitian version is rendered PT-symmetric by complexifying one of its coupling parameters that is responsible for an additional series of energy levels. This aspect of PCS potential was first pointed out by Bagchi and Quesne in \cite{bagchi00}, the second series of bound states showing up as resonances in its Hermitian version.\par
%
%
In a recent work, Abhinav and Panigrahi \cite{abhinav} have reiterated the well-known fact that to a large class of PT-symmetric complex potentials correspond two different superpotentials in the language of supersymmetric quantum mechanics. For the specific case of the PCS potential they have however missed an important point in their analysis, namely that the PCS potential exhibits an invariance under exchange of its coupling parameters that signals the appearance of two series of real eigenvalues for it. In the following, we will analyze Abhinav and Panigrahi's (AP) approach. Employing $\hbar = 2m = 1$ units, they considered the following general form of PCS potential
\begin{equation}
  V(x) = - V_1 \sech^2 \alpha x - {\rm i} V_2 \sech \alpha x \tanh \alpha x,
\end{equation}
together with the matching superpotentials
\begin{equation}
  W^{\pm}_{PT}(x) = (A \pm {\rm i} C^{PT}) \tanh \alpha x + (\pm C^{PT} + {\rm i} B) \sech \alpha x,
  \label{eq:W}
\end{equation}
where $A, B, C^{PT} \in \R$.\par
%
%
Using the notations of AP, corresponding to (\ref{eq:W}) the supersymmetric partner potentials can be read off from
\begin{equation}
  V_{\pm}(x) = W^2(x) \pm \frac{dW(x)}{dx}.
\end{equation}
In particular, for $V_-$ we have
\begin{equation}
\begin{split}
  V_-^{\pm}(x) & = - [(A \pm {\rm i} C^{PT}) (A \pm {\rm i} C^{PT} + \alpha) - (\pm C^{PT} + {\rm i} B)^2]
       \sech^2 \alpha x \\
  & \quad + (\pm C^{PT} + {\rm i} B) [2 (A \pm {\rm i} C_{PT}) + \alpha] \sech \alpha x \tanh \alpha x +
       (A \pm {\rm i} C_{PT})^2.  
\end{split}  \label{eq:V-}  
\end{equation}
This agrees with Eq.~(6) of AP except for the factorization energy term $(A \pm {\rm i} C_{PT})^2$. In the following arguments we suppress the latter except when necessary.\par
%
%
Rewriting the coefficients in (\ref{eq:V-}) by exposing the real and imaginary parts, we find
\begin{equation}
\begin{split}
  V_-^{\pm}(x) & = - [A^2 + B^2 - 2(C^{PT})^2 + \alpha A \pm {\rm i}(2A - 2B + \alpha) C^{PT}]
       \sech^2 \alpha x \\
  & \quad + \{\pm (2A - 2B + \alpha) C^{PT} + {\rm i} [2AB + 2(C^{PT})^2 + \alpha B]\} \sech \alpha x 
        \tanh \alpha x.  
\end{split}  \label{eq:V-bis}
\end{equation}
\par
%
%
It is evident from (\ref{eq:V-bis}) that in order for $V_-^{\pm}(x)$ to be PT-symmetric, one has to impose the unique constraint
\begin{equation}
  C^{PT} [2(A-B) + \alpha] = 0.  \label{eq:constraint}
\end{equation}
Of the two solutions provided by (\ref{eq:constraint}) only $C^{PT} = 0$ is non-trivial while the other, although examined by AP, only puts a relationship between the parameters $A$ and $B$, namely $A = B - \frac{\alpha}{2}$. We do not consider the latter possibility here.\par
%
%
{}For the solution $C^{PT} = 0$, it is readily implied by (\ref{eq:W}) that the associated superpotential is
\begin{equation}
  W^{\pm}_{PT}(x) = A \tanh \alpha x + {\rm i} B \sech \alpha x, \qquad A, B \in \R,  \label{eq:W-PT}
\end{equation}
leading to the following form for $V_-^{\pm}(x)$:
\begin{equation}
  V_-^{\pm}(x) = - [A (A + \alpha) + B^2] \sech^2 \alpha x + {\rm i} B  (2A + \alpha) \sech \alpha x 
  \tanh \alpha x - E,  \label{eq:V--PT}
\end{equation}
$E = - A^2$ being the factorization energy.\par
%
%
We stress that both Eqs.~(\ref{eq:W-PT}) and (\ref{eq:V--PT}) have been studied before by Bagchi and Quesne \cite{bagchi00}. That they observed for potential (\ref{eq:V--PT}) two series of real eigenvalues with physically acceptable wavefunctions is related to the fact that Eq.~(\ref{eq:V--PT}) is invariant under exchange of the parameters $A + \frac{\alpha}{2} \leftrightarrow B$, thereby leading to two possible candidates for the superpotentials
\begin{equation}
  W^{\pm}_{PT}(x) = A \tanh \alpha x + {\rm i} B \sech \alpha x, \qquad E = - A^2,
\end{equation}
\begin{equation}
  W^{\prime\pm}_{PT}(x) = \bigl(B - \tfrac{\alpha}{2}\bigr) \tanh \alpha x + {\rm i} \bigl(A + \tfrac{\alpha}{2}
  \bigr) \sech \alpha x, \qquad E' = - \bigl(B - \tfrac{\alpha}{2}\bigr)^2,
\end{equation}
with two different factorization energies $E$ and $E'$. The simultaneous existence of $W^{\pm}_{PT}$ and $W^{\prime\pm}_{PT}$ has been overlooked in \cite{abhinav} although it had already been pointed out in \cite{bagchi02a}.\par
%
%
To complete our discussion, let us also focus on the case $C^{PT} \ne 0$. Here we have at hand the form of $V_-^{\pm}(x)$ given by (\ref{eq:V-bis}). Its coefficient parameters are clearly complex. It is of interest to observe that by writing
\begin{equation}
  A \pm {\rm i} C^{PT} = {\cal A}, \qquad \pm C^{PT} + {\rm i} B = {\rm i} {\cal B},
\end{equation}
the potential $V_-^{\pm}(x)$ of (\ref{eq:V-bis}) can be exhibited in a form similar to (\ref{eq:V--PT}), namely 
\begin{equation}
  V_-^{\pm}(x) = - [{\cal A} ({\cal A} + \alpha) + {\cal B}^2] \sech^2 \alpha x + {\rm i} {\cal B}  (2{\cal A} 
  + \alpha) \sech \alpha x \tanh \alpha x - \varepsilon,  \label{eq:V--nonPT} 
\end{equation}
$\varepsilon = - {\cal A}^2$ being the factorization energy. Utilizing now the invariance under ${\cal A} + \frac{\alpha}{2} \leftrightarrow {\cal B}$, we have for (\ref{eq:V--nonPT}) two possible candidates for the superpotentials
\begin{equation}
  {\cal W}^{\pm}_{PT}(x) = {\cal A} \tanh \alpha x + {\rm i} {\cal B} \sech \alpha x, \qquad \varepsilon = 
  - {\cal A}^2,
\end{equation}
\begin{equation}
  {\cal W}^{\prime\pm}_{PT}(x) = \bigl({\cal B} - \tfrac{\alpha}{2}\bigr) \tanh \alpha x + {\rm i} \bigl({\cal A} + 
  \tfrac{\alpha}{2}\bigr) \sech \alpha x, \qquad \varepsilon' = - \bigl({\cal B} - \tfrac{\alpha}{2}\bigr)^2.
\end{equation}
Note that $\varepsilon, \varepsilon' \in \C$.\par
%
%
{}Finally we can compare the general complexified form (\ref{eq:V-bis}) of $V_-^{\pm}(x)$ with the $sl(2)$ family defined by \cite{bagchi00, bagchi02a}
\begin{equation}
  V_m = \bigl(\tfrac{1}{4} - m^2\bigr) F' + 2m G' + G^2,
\end{equation}
where $m = m_R + {\rm i} m_I$ and $F$, $G$ are given by $F = \tanh \alpha x$, $G = (b_R + {\rm i} b_I) \sech \alpha x$ with $m_R, m_I, b_R, b_I \in \R$. The potential $V_m$ turns out to be
\begin{equation}
\begin{split}
  V_m & = [b_R^2 - b_I^2 - \alpha (m_R^2 - m_I^2) + \tfrac{1}{4} \alpha + {\rm i} (2b_R b_I - 2 \alpha m_R
       m_I)] \sech^2 \alpha x \\
  & \quad - 2\alpha [m_R b_R - m_I b_I + {\rm i} (m_R b_I + m_I b_R)] \sech \alpha x \tanh \alpha x. 
\end{split}
\end{equation}
\par
%
%
Comparison with (\ref{eq:V-bis}) produces the following correspondences
\begin{equation}
\begin{split}
  & b_R^2 - b_I^2 - \alpha(m_R^2 - m_I^2) + \tfrac{1}{4} \alpha = - [A^2 + B^2 - 2(C^{PT})^2 + \alpha A], \\
  & 2b_R b_I - 2\alpha m_R m_I = \mp (2A - 2B + \alpha) C^{PT}, \\
  & - 2\alpha (m_R b_R - m_I b_I) = \pm (2A - 2B + \alpha) C^{PT}, \\
  & - 2\alpha (m_R b_I + m_I b_R) = 2AB + 2(C^{PT})^2 + \alpha B.  \label{eq:sl-susy}
\end{split}
\end{equation}
The last two equations of (\ref{eq:sl-susy}) can be solved to yield
\begin{equation}
\begin{split}
  & m_R = \frac{1}{\alpha (b_R^2 + b_I^2)} \left\{\mp b_R \left(A - B + \frac{\alpha}{2}\right) C^{PT}
       - b_I \left[AB + (C^{PT})^2 + \frac{\alpha}{2} B\right]\right\}, \\
  & m_I = \frac{1}{\alpha (b_R^2 + b_I^2)} \left\{\pm b_I \left(A - B + \frac{\alpha}{2}\right) C^{PT}
       - b_R \left[AB + (C^{PT})^2 + \frac{\alpha}{2} B\right]\right\},
\end{split}
\end{equation}
which imply
\begin{equation}
\begin{split}
  m_R^2 - m_I^2 & = \frac{1}{\alpha^2 (b_R^2 + b_I^2)^2} \Bigl\{(b_R^2 - b_I^2) \Bigl[(C^{PT})^2 \Bigl(
       A - B + \frac{\alpha}{2}\Bigr)^2 \\
  & \quad - \Bigl(AB + (C^{PT})^2 + \frac{\alpha}{2} B\Bigr)^2\Bigr] \\
  & \quad \pm 4b_R b_I C^{PT} \Bigl(A - B + \frac{\alpha}{2}\Bigr) \Bigl(AB + (C^{PT})^2 + \frac{\alpha}{2} B
       \Bigr)\Bigr\} 
\end{split}
\end{equation}
and
\begin{equation}
\begin{split}
  m_R m_I & = \frac{1}{\alpha^2 (b_R^2 + b_I^2)^2} \Bigl\{\pm C^{PT} (b_R^2 - b_I^2) \Bigl(A - B + 
       \frac{\alpha}{2}\Bigr) \Bigl(AB + (C^{PT})^2 + \frac{\alpha}{2} B\Bigr) \\
  & \quad - b_R b_I \Bigl[(C^{PT})^2 \Bigl(A - B + \frac{\alpha}{2}\Bigr)^2 - \Bigl(AB + (C^{PT})^2 + 
       \frac{\alpha}{2} B\Bigr)^2\Bigr]\Bigr\}.  
\end{split}
\end{equation}
\par
%
%
Knowing $m_R$ and $m_I$ gives the corresponding solutions of $b_R$ and $b_I$ from the first two equations of (\ref{eq:sl-susy}). This confirms the point that the formulae governing $sl(2)$ for $m \in \C$ are completely equivalent to the starting relations of AP.\par
%
%
In conclusion, we have shown that having missed an essential invariance property of the PCS potential under parameter exchange, AP failed to notice the existence of a second series of real eigenvalues whenever the condition $C^{PT} = 0$ is satisfied. As previously remarked, there are two associated PT-antisymmetric superpotentials in such a case and not only one as AP claim. In addition, PT-symmetry breaking is produced by the transition from $C^{PT} = 0$ to $C^{PT} \ne 0$ and not by the appearance of a subtle relation between the parameters $A$ and $B$. In such a process, the two PT-antisymmetric superpotentials go into two non-PT-antisymmetric ones connected with the pairs of complex conjugate energy eigenvalues.\par
%
%
Although we have analyzed here only the case of the PCS potential, similar considerations apply to some other potentials too \cite{bagchi00, bagchi02a, bagchi01, bagchi02b}. Finally it is worth stressing that our observations have been confirmed by some independent studies \cite{levai01, levai02}.\par
%
%


\newpage
\begin{thebibliography}{99}

\bibitem{bender98} C.M.\ Bender, S.\ Boettcher, Phys.\ Rev.\ Lett.\ 80 (1998) 5243.

\bibitem{bender07} C.M.\ Bender, Rep.\ Prog.\ Phys.\ 70 (2007) 947.

\bibitem{bagchi00} B.\ Bagchi, C.\ Quesne, Phys.\ Lett.\ A 273 (2000) 285.

\bibitem{bagchi02a} B.\ Bagchi, C.\ Quesne, Phys.\ Lett.\ A 300 (2002) 18.

\bibitem{bagchi01} B.\ Bagchi, S.\ Mallik, C.\ Quesne, Int.\ J.\ Mod.\ Phys.\ A 16 (2001) 2859.

\bibitem{bagchi02b} B.\ Bagchi, S.\ Mallik, C.\ Quesne, Int.\ J.\ Mod.\ Phys.\ A 17 (2002) 51.

\bibitem{abhinav} K.\ Abhinav, P.K.\ Panigrahi, Ann.\ Phys.\ 325 (2010) 1198.

\bibitem{levai01} G.\ L\'evai, F.\ Cannata, A.\ Ventura, J.\ Phys.\ A: Math.\ Gen.\ 34 (2001) 839.

\bibitem{levai02} G.\ L\'evai, M.\ Znojil, J.\ Phys.\ A: Math.\ Gen.\ 35 (2002) 8793.
 
\end{thebibliography}
\end{document}